\documentclass[aps,prl,reprint]{revtex4-1}
\usepackage{graphicx}
\usepackage{dcolumn}
\usepackage{bm}
\usepackage{color}
\usepackage{siunitx}
\usepackage{amsmath}
\usepackage{amssymb}
\usepackage{textgreek}

\begin{document}

\preprint{APS/123-QED}

\title{A two-way photonic interface for linking Sr$^+$ transition at 422\,nm to the telecommunications C-band}

\author{Thomas A. Wright\textsuperscript{1}}
\email{t.wright@bath.ac.uk}
\author{Robert J.A. Francis-Jones\textsuperscript{1,2}}%
\author{Corin B.E. Gawith\textsuperscript{3}}%
\author{Jonas N. Becker\textsuperscript{2}} 
\author{Patrick M. Ledingham\textsuperscript{2}}%
\author{Peter G.R. Smith\textsuperscript{3}}
\author{Joshua Nunn\textsuperscript{1}}%
\author{Peter J. Mosley\textsuperscript{1}}%
\author{Benjamin Brecht\textsuperscript{2}}%
\author{Ian A. Walmsley\textsuperscript{2}}%

\affiliation{%
\textsuperscript{1}Centre for Photonics and Photonic Materials, Department of Physics, University of Bath, Bath, BA2 7AY, UK
}%
\affiliation{%
\textsuperscript{2}Clarendon Laboratory, University of Oxford, Parks Road, Oxford, OX1 3PU, UK
}%
\affiliation{%
\textsuperscript{3}Optoelectronics Research Centre, University of Southampton, Southampton, SO17 1BJ, UK
}%


\date{\today}

\begin{abstract}
We report a single-stage bi-directional interface capable of linking $\text{Sr}^{\text{+}}$ trapped ion qubits in a long-distance quantum network. Our interface converts photons between the $\text{Sr}^{\text{+}}$ emission wavelength at 422\,nm and the telecoms C-band to enable low-loss transmission over optical fiber. We have achieved both up- and down-conversion at the single photon level with efficiencies of 9.4\,$\%$ and 1.1\,$\%$ respectively. Furthermore we demonstrate noise levels that are low enough to allow for genuine quantum operation in the future.
\end{abstract}

\maketitle

\section{Introduction}
Large scale quantum networks suitable for long-distance secure communication and distributed computation require not only that quantum information may be manipulated reliably, but also communicated successfully between remote nodes \cite{Kimble:2008if,Anonymous:2001dt}. However, existing quantum information processing platforms are not individually able to fulfil both of these requirements. For example, photons can distribute quantum information through fiber networks \cite{Sun:2017bz} or via satellite \cite{Yin:2017im} but multi-photon gates remain challenging, whereas trapped ions have achieved high-fidelity two-qubit operations \cite{Ballance:2016hy} but are unsuitable for sharing entanglement beyond a single laboratory. However, integrating disparate technologies to form a hybrid light-matter quantum network promises the capability to carry out entanglement distribution and quantum communication over long distances \cite{Walmsley:2016hn,Nigmatullin:2016dx}.

The optical communication bus between nodes must overcome two technical challenges: compatibility between devices operating at different optical frequencies and low loss across large separations. Quantum frequency conversion (QFC), where a photon is coherently shifted to a different frequency band, addresses this difficulty by linking wavelengths as short as the ultraviolet (UV), where many convenient ion transitions are located, and the infrared (IR) telecommunications bands, enabling long-distance low-loss transmission in optical fiber \cite{Kumar:1990km}. 

A diverse range of platforms have been utilized to demonstrate frequency conversion via three- or four-wave mixing, including nonlinear crystals \cite{Albota:2004gm,Samblowski:2014in}, planar waveguides \cite{Langrock:2005ie,Tanzilli:2005jd}, optical fibers \cite{McGuinness:2010ja,Clark:2013cia}, microresonators \cite{Li:2016eg,Guo:2016iu} and atomic systems \cite{Radnaev:2010il,Bustard:2017dh}. QFC experiments initially focused on enhancing detection of IR photons by mapping them to the visible and near-infrared (NIR) where efficient silicon photon detectors existed \cite{Albota:2004gm,Vandevender:2004ef}. QFC has since been shown to achieve the frequency remapping of both squeezed light \cite{Vollmer:2014ia,Kong:2014it} and entangled states \cite{Tanzilli:2005jd,Ikuta:2011de}. The enhancement offered by QFC to optical interconnects within a quantum network has led to a demonstration of direct coupling between dissimilar quantum memories \cite{Maring:2017eb}, as well as  a multitude of experiments translating node-compatible photons both from \cite{Tanzilli:2005jd,Vandevender:2007gx,Kamada:2007gf,Rakher:2010jua,Rakher:2011jia,Maring:2014dp,Vollmer:2014ia,Baune:2016fd,Allgaier:2017hv} and to \cite{Takesue:2010ba,Ikuta:2011de,Zaske:2012hka,DeGreve:ge,Forchel:2012ka, Albrecht:cf,Farrera:2016jt,Krutyanskiy:2017eu,Bock:2017tr,Walker:2017vd,Dreau:2018vs} different telecommunication bands. 

Near-unit efficiency frequency conversion has been demonstrated \cite{Clark:2013cia}, although, in practice, complexities arising from photon bandwidth, pump induced noise and the required frequency shift have limited the capability of many converters, proving arduous challenges to overcome. QFC via three-wave mixing, in particular, imposes the strict requirement that the high nonlinear coupling be provided by a singular strong pump which additionally must account for the energy difference between the input and target wavelength. In large part, existing QFC experiments have exploited opportune laser wavelengths and transitions predominantly in the red and NIR. The large frequency separation between IR and and the blue - UV makes connecting most ion traps to the telecom bands particularly challenging, leading to the proposition of a two stage approach \cite{Clark:2011hp}.  Classical generation of UV light by means of parametric three-wave mixing is in itself difficult, so far being realized by SHG \cite{Wang:1998gc} and sum frequency generation (SFG) \cite{Oh:1995gr}. Recently however, developments in the translation of quantum states between ultraviolet and the O-band has been shown \cite{Rutz:2016ej,PhysRevApplied.7.024021}, albeit only in one direction. The majority of reported conversions thus far have been unidirectional with limited exceptions \cite{Dudin:2010ji}, whereas to create a functional quantum network in fiber, shifting to the telecoms is critical and two-way conversion is desirable.

We report the realization of single-stage bi-directional frequency conversion at the single photon level for interfacing the \(\text{S}_{1/2} \rightarrow \text{P}_{1/2}\) transition in trapped $\text{Sr}^{\text{+}}$ ion qubits (422\,nm) with the telecommunication C-band (1550\,nm). The conversion is achieved in a magnesium-doped periodically poled lithium niobate (MgO:PPLN) crystal, where \( \chi^{(2)}\) sum or difference frequency generation (SFG or DFG) can be used to achieve up or down conversion of an input photon. By monitoring the conversion of weak coherent light with single-photon detectors, we demonstrate that the noise level, expressed as the actual noise photons normalized to the conversion efficiency, $\mu_1$ \cite{Gundogan:2015dh}, is as low as 0.0185 -- far below the level required for use as the interface in a hybrid quantum network. This is to our knowledge the only bi-directional link between blue ion transitions and the C-band.

\section{Technical Details}
In order to map the input to the target output wavelength, a strong pump field must be tuned to fulfill the energy conservation requirement $\hbar\omega_{\text{in}} + \hbar\omega_{\text{pump}} = \hbar\omega_{\text{out}} $ for SFG up conversion and $\hbar\omega_{\text{in}} - \hbar\omega_{\text{pump}} = \hbar\omega_{\text{out}} $ for DFG down conversion. In both SFG and DFG the amplitude of the strong pump serves to drive the nonlinear optical response facilitating the conversion. The efficiency of the process is related to the strength of the coupling between the fields in the QFC Hamiltonian:
\begin{equation}
\label{eq.1}
\hat{\mathcal{H}} = i\hbar\kappa A_{\text{pump}} \big( \hat{a}^\dagger_{\text{out}} \hat{a}^{}_{\text{in}} - h.c.\big)  
\end{equation}
determined by the parameter \(\kappa\), which itself is dependent on the magnitude of the fields but also on their relative phase, spatial overlap in the crystal and a the intrinsic nonlinearity of the material \cite{Kumar:1990km}. 

For such widely separated wavelengths, the large wave-vector mismatch \(\Delta k \) between the propagation constants of the three fields means that phase matching is difficult to satisfy in commonly-available materials. Even typical quasi phase-matched (QPM) crystals that achieve efficient conversion when 
\begin{equation}
\label{eq.2}
\Delta k - \frac{2 \pi}{\Lambda} = k _{\text{out}} - k _{\text{in}} - k _{\text{pump}} - \frac{2 \pi}{\Lambda} =0
\end{equation} 
have a poling pitch \(\Lambda\) that is too long to compensate the large \(\Delta k \) in our interaction. Hence, we used a MgO:PPLN crystal fabricated in collaboration with Covesion Ltd with ferroelectric domains created using a proprietary electric field poling technique to produce a very short pitch, shown in Fig.\,1. A photoresist pattern was created on the -$z$ (bottom) face of a 0.5 mm-thick single-domain $z$-cut 3 inch diameter MgO:LiNbO$_3$ crystal wafer. Liquid electrodes were applied to both the patterned -$z$ and unpatterned +$z$ surfaces of the crystal to enable electrical contact with the wafer surface. Domain inversion along the $z$-axis was performed at room temperature by voltage-controlled application of electric field based on a first stage of domain nucleation above the coercive field of the crystal, and a second stage of domain spreading near the coercive field  ($\sim$\SI{4.5}{\kilo\volt\per\milli\metre}); this technique results in inverted domains that traverse the entire thickness of the crystal. The MgO:PPLN wafer was diced and polished into multiple chips, each containing five \SI{300}{\micro\meter}-wide gratings with periods of 3.75, 3.85, 3.95, 4.05, and \SI{4.15}{\micro\meter} respectively. These periods were calculated to enable SFG and DFG processes between 422\,nm and the telecoms C-band. The MgO:PPLN crystal used in this experiment was 19.97\,mm long, with the \SI{3.75}{\micro\meter} grating selected; it was not anti-reflection coated.

\begin{figure}
	\centering
 	\includegraphics[width=0.48\textwidth]{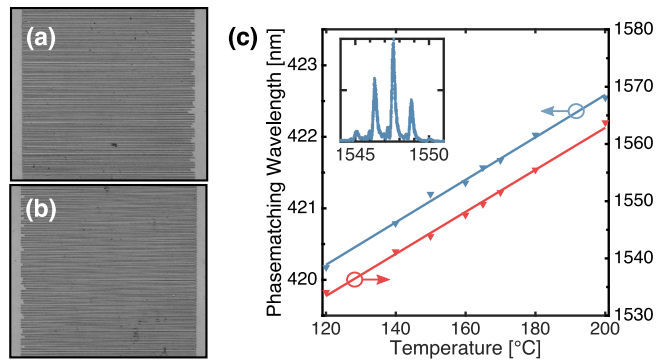}
	\caption{Optical micrographs of the (a) +$z$ (top) and (b) -$z$ (bottom) surface of one of the \SI{3.75}{\micro\meter} periodically poled MgO:LiNbO3 crystals. (c) Phase-matching characteristics. The temperature response of the phase-matching of the \SI{3.75}{\micro\meter} poled crystal, i.e., the wavelength of perfect phase matching as a function of crystal temperature. \textit{Inset} SFG phase-matching curve, ie., short wavelength output power as a function of input telecom wavelength at constant crystal temperature and pump power.}
\end{figure}

\section{Experimental Demonstration}
\subsection{Phase-matching characterization}

The experimental setup for characterizing the SFG up-conversion is shown in Fig. 2(a). An 80\,MHz synchronously pumped tunable dye laser operating at 580\,nm wavelength and 30\,ps pulse duration (Sirah Gropius) was used to pump the conversion and a tunable continuous-wave (CW) laser with a 40\,MHz linewidth (Santec  TSL-510 C), provided a coherent IR input that was attenuated to low mean photon number. The pump and input beam size and polarization were set using telescopes, polarizing beam splitters (PBSs), and half-wave plates (HWPs) before the beams were combined at a dichroic mirror (DM) and directed towards the MgO:PPLN crystal. A pair of fused silica lenses were then used to focus the overlapped beams into the crystal. Care was taken to balance matching the Rayleigh length of each beam to half the crystal length while minimizing the difference between the cross-sections of the beam waists. Following the crystal, a flipper mirror allowed for the input and pump powers transmitted through the crystal to be measured. The pump, unconverted IR input light and successfully converted violet output were then separated using a series of dichroic and short pass (SPF) filters before being directed to two single-photon avalanche diodes (SPADs). At the short wavelength, the pump light was removed using short pass filters at 500\,nm and 440\,nm as well as a short-pass dichroic filter with an edge at 500\,nm. The remaining signal was monitored by a blue-enhanced Si SPAD with a detection efficiency, dead time and dark count rate of 86\,\%, 43\,ns and 6\,Hz. For the long wavelength we utilized a fiber-coupled InGaAs SPAD operating at 9.5\,\% detection efficiency with a dead time of \SI{10}{\micro\second}. A time-tagging module (TTM) was used to record counts from the detectors. A spectrograph with electron-multiplying CCD camera was available to monitor the spectra of the violet light. 

In order to maximize the coupling constant, \(\kappa\), the phase matching of the crystal was first characterized. With the temperature of the crystal stabilized at 160\(^\circ\)C, the input IR beam was swept in wavelength whilst the intensity of the converted violet light was measured, mapping out the phase matching curve, for a pump wavelength of 579.6\,nm. We observed several distinct peaks in the phase-matching (see Fig. 1), indicative of either multiple frequency modes within the pump beam or inhomogeneous poling across the length of the crystal.

We measured the change in position of the central phase matching peak as a function of temperature, for a range of crystal temperatures by sweeping the input wavelength whilst measuring the output violet power using an amplified photodiode. For the input IR light we measured a temperature response of \(\Delta \lambda _{in} / \Delta T\) = 0.4\,nm/K, corresponding to a change in the output wavelength of \(\Delta \lambda _{\text{out}} / \Delta T\) =  0.0297\,nm/K.

\begin{figure}
	\centering
 	\includegraphics[width=0.48\textwidth]{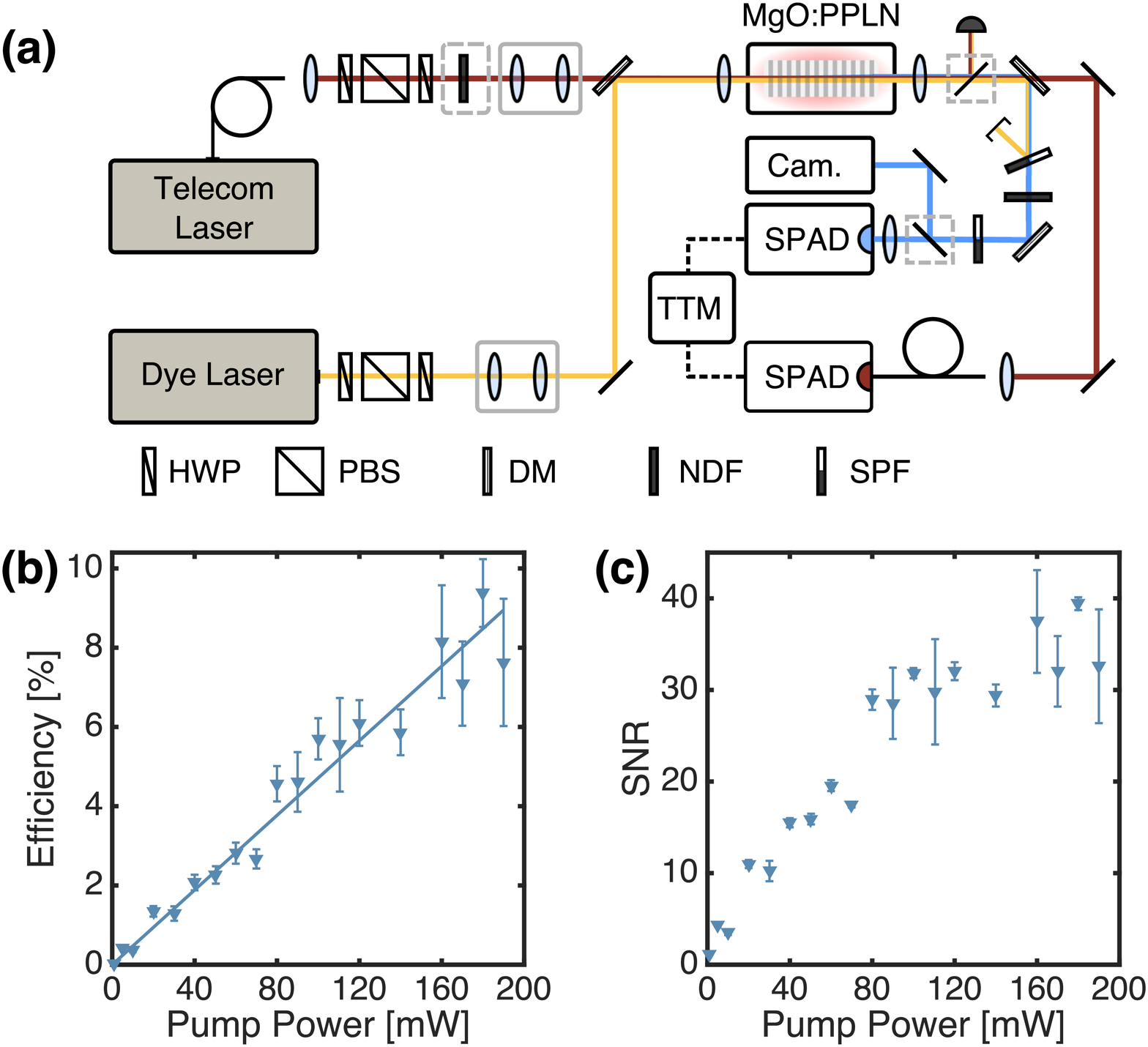}
	\caption{(a) Experimental setup of the SFG up conversion. (b) Up conversion efficiency of the interface. (c) Signal-to-noise ratio of the up conversion.}
\end{figure}

\subsection{Frequency up-conversion}
Having established the phase matching response of the crystal, we investigated the achievable up conversion efficiency. IR light at 1547.6\,nm was converted to 421.7\,nm by pumping the process at 579.6\,nm with a fixed crystal oven temperature of 160\(^\circ\)C. 

We present the external efficiency, \(\eta _{\text{ext}}\), of the SFG conversion in Fig. 2(b), which we define as the mean number of converted photons per second leaving the crystal divided by the number of input photons per second incident and temporally overlapped with the pump. The number of input photons per second, $\langle n\rangle_{in}$, is therefore given by 
\begin{equation}
\label{eq.4}
\langle n\rangle _{\text{in}} = \frac{P_{\text{in}} D}{\hbar \, \omega_{\text{in}}},
\end{equation} 
where \(P_{\text{in}}\) was the IR power transmitted through the crystal measured with the pump beam blocked. Due to observations of drift in the IR input power, this was measured both before and after each integration over which we recorded detector counts and used to calibrate the continuous monitoring of the IR light at SPAD detector 2. Due to the CW nature of the input light, the duty cycle $D$ was defined as the pump pulse width multiplied by the repetition rate: \( D = \tau _{\text{pump}} \cdot R_{\text{p}} \). 

The conversion efficiency was then calculated as
\begin{equation}
\label{eq.3}
\eta _{\text{ext}} = \frac{S-N}{\langle n \rangle _{\text{in}}\, \eta _{\text{\,loss}}},
\end{equation}
where \( \eta _{\text{loss}} \) incorporates the detector efficiency (86\,$\%$), transmission through optical components (96\,$\%$) and, at higher powers, a neutral density filter (NDF) with approximately 20\,$\%$ transmission used to prevent detector saturation. $S$ and $N$ are the signal and noise counts per second respectively. As the SPAD detectors output only a single click when one or more photons arrive within their dead time, $S$ and $N$ were corrected relative to the measured rates $S_{\text{raw}}$ and $N_{\text{raw}}$ as follows: 
\begin{equation}
\label{eq.5}
S = \frac{S_{\text{raw}}}{1-S_{\text{raw}} T_{\text{D}}}; \hspace{5mm}
N = \frac{N_{\text{raw}}}{1-N_{\text{raw}} T_{\text{D}}},
\end{equation}
where $T_{\text{D}} \gg 1/R_{\text{p}}$ is the dead time of the detector. The background noise was measured by blocking the IR input and recording counts at the visible detector with the pump unblocked.

In order to demonstrate the capability of the interface to operate at low photon numbers we set a target input of overlapping an average of two IR photons from the IR attenuated coherent source with every pump pulse. The resulting conversion efficiency is shown in Figure 2(b). The highest measured external efficiency of \( \eta _{ext} \) = 9.4$\pm$0.86\,$\%$ was observed for an average pump power of 180\,mW. Each point in Figure 2(b) corresponds to a measurement consisting of between 5 and 10\,s of integration, beyond which we observed a drop in efficiency. This was later realized to be due to absorption of the pump light causing localized heating of the crystal, resulting in a change in the phase matching. The phenomenon of pump-power-induced change in \(\Delta k\) for \(\chi^{(2)}\) QFC processes has been discussed previously \cite{Louchev:2005ij,PhysRevApplied.7.024021}. Additional change in the phase mismatch is also introduced due to the photorefractive effect and as such, when operating over extended periods of time as would be required in a network, the process would be pumped at a constant power, with the phase matching temperature tuning being optimized at the selected pump power. Figure 2(c) shows the signal to noise ratio (SNR) achieved across a range of pump powers, where we define SNR = $S/N$. The SNR achieved for the point at which we achieved highest conversion was 39.4$\pm$0.69.

\subsection{Frequency down-conversion}

In order to demonstrate a two way interface, we similarly characterized the reverse process, converting single-photon level violet light to IR via DFG. Figure 3(a) shows the modified experimental setup. The input light at 425.5\,nm was obtained by second harmonic generation of a 80\,MHz repetition rate Ti:sapphire laser operating at 851\,nm wavelength and 300\,ps pulse duration (Spectra Physics Tsunami) which was synchronized to the clock signal of the dye laser system via active cavity-length control. Replicating the interface between 421.7\,nm and the C-band was not possible due to the phase matching restriction of the SHG crystal. In order to successfully translate the violet light to the telecoms C band we tuned the wavelength of the dye laser to 585\,nm and adjusted the crystal oven temperature to 226.4\(^\circ\)C. This enabled us to optimize conversion to 1560.6\,nm. Mitigation of the pump induced change in \(\Delta k \) was achieved by optimizing the oven temperature whilst pumping the nonlinear conversion with an average power of 60\,mW, half of the available range. In the long wavelength detection arm the filtering consisted of two long pass filters with edges at 950 and 650\,nm. A 8.9\,nm-wide band pass filter (BPF) was used, centered at 1570\,nm and rotated in order to shift the transmission to accommodate the converted light at 1560\,nm.

\begin{figure}[ht]
	\centering
 	\includegraphics[width=0.48\textwidth]{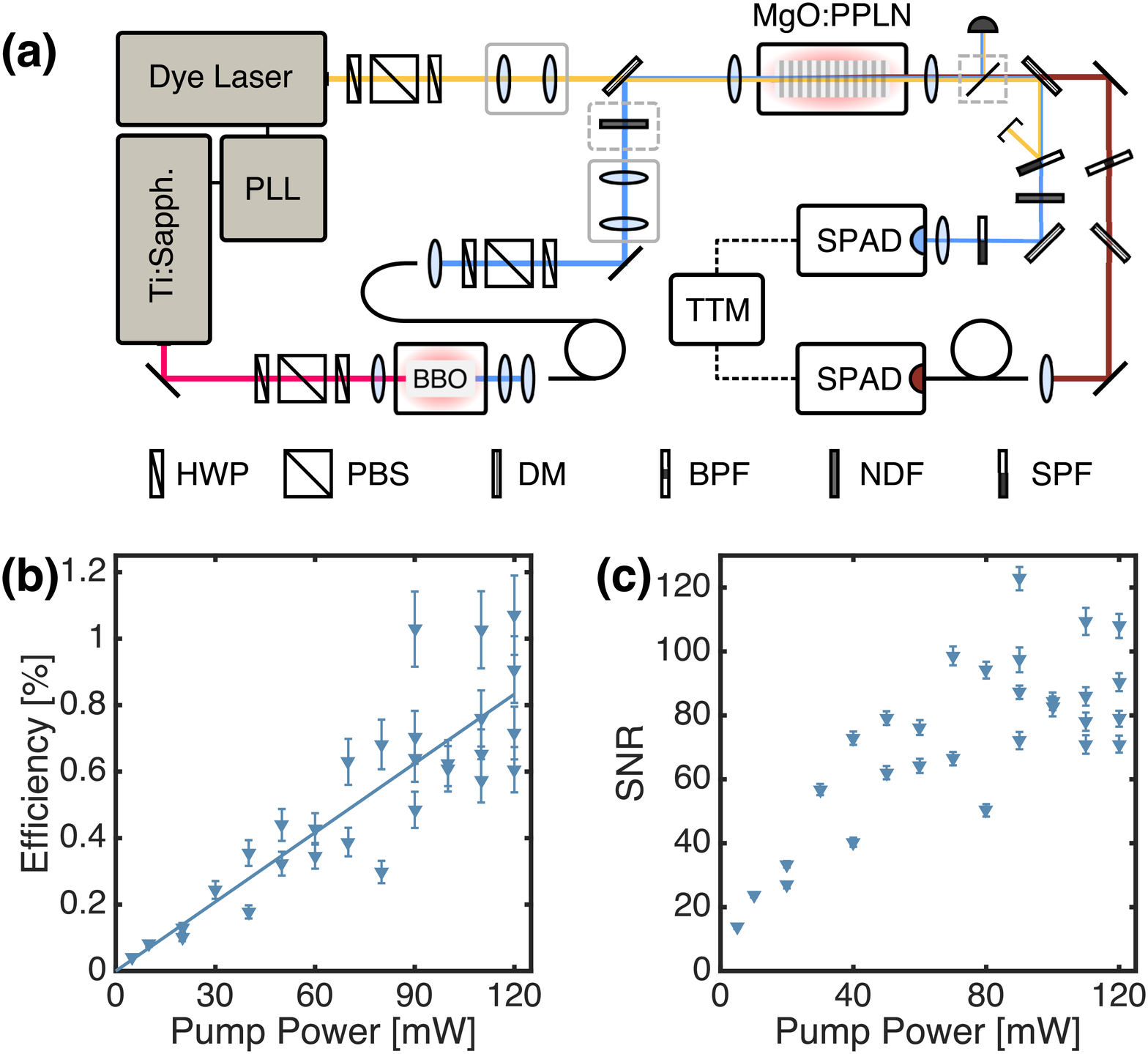}
	\caption{(a) Experimental setup of the DFG down conversion. PLL: phase-locked loop (b) Down-conversion efficiency of the interface. (c) Signal-to-noise ratio of the down conversion.}
\end{figure}

In Fig. 3(b) we present the external efficiency of the conversion, where we have accounted for sources of loss outside the crystal, as described by Eq. 3. We 
evaluated \( \eta _{loss} \), incorporating the detector efficiency ($\sim$ 9.5\,$\%$), transmission through optical components ($\sim$ 73\,$\%$) and fiber coupling efficiency ($\sim$ 65\,$\%$). The duty cycle D was defined as the ratio of the pump and input pulse durations: \(D = \tau _{pump} / \tau _{input} \). The same target input photon number as the SFG conversion, of $\langle n \rangle _{in} =$ 2 per pulse, was used in the DFG experiment.  Across the data collected, counts were measured for integration times longer than 15\,s. 

The pump and input beam were steered using mirrors ahead of the crystal to optimize the beam overlap leading to a wide distribution of observed conversion at comparable pump powers. A maximum external conversion efficiency of (1.1\,$\pm$ 0.12\,)$\%$ was achieved for a pump power of 120\,mW, While significantly lower than the $\sim$ 6\,$\%$ external efficiency observed for the SFG at equal pump power, the observed value may be partly explained by a change in beam waist between the two experiments. The beam waist of the pump in both experiments was $\sim$ \SI{43.2}{\micro\meter}, while the input light beam waist was increased from $\sim$ \SI{63.3}{\micro\meter} in the SFG up conversion to $\sim$ \SI{112}{\micro\meter} in the DFG down conversion. This would indicate that the maximum proportion of the input overlapped by the pump changed from $\sim$ 46.6\,$\%$ to $\sim$ 14.9\,$\%$. It could be expected that further optimization of optics selection may yield improved conversion efficiencies, in particular when considered for the DFG down-conversion. Again, we draw attention to the conversion efficiency being limited by the available pump power in the experiment, as can be seen from the linear slope of the efficiency vs pump power curve in Fig. 3(b). The down conversion was shown to be low noise in operation, with a SNR of 108$\pm$3.8 measured at the point of highest conversion.

\section{Discussion}
\subsection{Noise analysis}
In order to demonstrate that our device is capable of operating as a QFC interface in a quantum network, for both up and down conversion, we calculated the mean input photon number per pulse that would yield a SNR = 1, commonly referred to as $ \mu _1$ \cite{Gundogan:2015dh}. Originally defined for quantum memories, the parameter $ \mu _1$ provides a useful performance benchmark for frequency conversion as it normalizes the noise by the conversion efficiency, thus precluding noise reduction by pumping with unrealistically low power. The calculated values are shown in Figure 4(a). For the up conversion we found the minimum $\mu _1$ = 0.05074 when operating at a pump power of 180\,mW. We note that this value was limited by the pump power in the experiment, as can be seen from the linear slope of the efficiency vs pump power curve, and that these values are obtained without narrow-band spectral filtering. For the down conversion, we find $\mu_1$ =  0.0185 at a pump power of 120\,mW. 

\begin{figure}
	\centering
 	\includegraphics[width=0.48\textwidth]{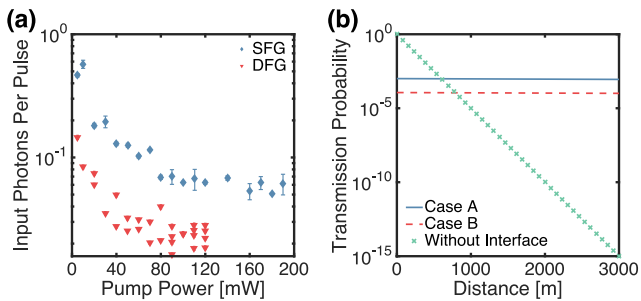}
	\caption{(a) \(\mu _1\), The lowest number of photons per pulse possible to send into the device to get an SNR \(\geq\) 1. (b)  Probability of successful transmission of two photons from remote trapped Sr$^+$ ion processing nodes to an entangler along a fiber network with and without our interface.}
\end{figure}

The very low value of $\mu_1$ for down conversion shows that noise originating from broadband, un-phase-matched spontaneous parametric down conversion (SPDC) of the pump field -- typically a limiting factor in $\chi^\text{(2)}$ QFC schemes linking short to long wavelengths \cite{Pelc:2011eh} -- does not inhibit the capability of our system to operate at the single-photon level. As a result, we believe that up converted SPDC, which has been shown to contribute broadband noise to frequency conversion \cite{Fix:2018et,PhysRevApplied.7.024021}, is not the dominant noise source in our SFG configuration. We attribute the contrast in SNR between down conversion and up conversion primarily to the difference in detector sensitivity at the pump wavelength: the Si visible detector is sensitive to the pump wavelength whereas the InGaAs IR detector is not. The resulting inability to detect pump light leaking through the filters during down conversion and commensurate lower noise level suggests that pump leakage is the dominant source of noise in the up conversion configuration. This could in principle be removed by increasing the optical depth of the filters at the pump wavelength.

Our values of $\mu_1 \ll 1$ demonstrate that low-noise operation is possible without narrow-band spectral filtering. Although high-efficiency conversion of nanosecond-duration photons from Sr$^+$ ions would require increased pulse energy, due to the broadband nature of the noise and its linear dependence on pulse energy we expect that low-noise conversion could be maintained through narrow band filtering as seen in \cite{Farrera:2016jt,Walker:2017vd}. To illustrate, we consider our device operation in ion compatible down-conversion, utilizing 1\SI{}{\micro\second} duration pump pulses. In our current configuration we observe 3.9$\times \text{10}^{\text{-6}}$ noise photons per 300\,ps pump pulse, at the point of highest conversion. Increasing pulse duration to \SI{1}{\micro\second} would see this number increase by a factor of $\sim$ 3300. However, replacing our existing 8.9\,nm band pass filter at 1570\,nm with a 200\,MHz filter, similar to that used in \cite{Farrera:2016jt} would see a reduction of broadband noise by a factor of $\sim$ 5400, hence maintaining a similar noise contribution overall.

\subsection{Projected network enhancement}

In Figure 4(b) we consider the end-to-end efficiency of three scenarios in which entanglement needs to be established between remote processing nodes, taking into account the transmission loss of fiber at the wavelengths 422\,nm and 1550\,nm ($<$50\,dB/km for SM400 at 422\,nm and $<$0.18\,dB/km for SMF-28 at 1550\,nm).  In case A a violet photon from node 1 is down-converted via our interface and transmitted through a length of optical fiber before reaching the location of the second node, where it is up-converted to allow interference with a photon emitted directly from the node 2. In case B both photons emitted by the ion traps are down-converted before being transmitted through optical fiber and interfered at some midway position. The final case is where no interface is used and a violet photon is emitted from one of the nodes before being transmitted through fiber to the second node whereupon an entanglement link is established. For each case we present the probability of the photons successfully reaching their destination, assuming no coupling loss. We see that the use of our interface would drastically increase the probability of successful end-to-end transmission for remote nodes linked by fiber, even for typical intra-city distances of a few kilometres.

\section{Conclusion}
We have implemented an interface capable of low noise up- and down-conversion of single-photon-level light with efficiencies of 9.4$\,\%$ and 1.1\,$\%$ respectively in a custom-poled MgO:PPLN crystal. When considering the transmission loss of single mode fiber at the  Sr$^+$ emission wavelength of 422\,nm relative to that at 1550\,nm, our interface would increase the probability of successful transmission of quantum information by 47 orders of magnitude over a distance of 10\,km. We have demonstrated that the noise introduced when converting a weak coherent state is far below the level required to achieve high-fidelity operation with single photons emitted by trapped Sr$^+$ ions. Hence we believe that our interface will enable long-distance entanglement distribution through chains of nodes containing trapped Sr$^+$ ions, paving the way to the construction of large-scale quantum networks. 

\section{Acknowledgements}
This work was funded by the UK EPSRC Quantum Technology Hub Networked Quantum Information Technologies, grant number EP/M013243/1. 
B.B. acknowledges funding  from the European Union (EU) Horizon
2020 Research and Innovation Program under Grant No. 665148. P.M.L. acknowledges funding from  the European Union Horizon 2020 Research and Innovation Framework Programme Marie Curie individual fellowship, Grant No. 705278.

\bibliography{bibliography.bib}

\end{document}